\RequirePackage{amsmath}
\documentclass[12pt]{iopart}
\usepackage{graphicx}
\usepackage{color}
\usepackage{mathtools}
\usepackage[normalem]{ulem}
\newcommand{\Bbbk}{\mathbf{k}}
\def\clr{\color{black}}

\newcommand {\m}{{\kappa}}

\begin{document}
\title{Phase separation transition  of   reconstituting $k$-mers in one dimension}
\author{Bijoy  Daga$^{1}$ and P. K. Mohanty$^{1,2}$}
\ead{ bijoy.daga@saha.ac.in, pk.mohanty@saha.ac.in} 
\address{  
 $^{1}$Condensed Matter Physics Division, Saha Institute of Nuclear Physics, 1/AF Bidhan Nagar, Kolkata 700064, India\\
 $^{2}$Max Planck Institute for the Physics of Complex Systems, N\"othnitzer Stra\ss e 38, 01187 Dresden, Germany
 }

\begin{abstract}
We introduce a driven diffusive model involving poly-dispersed hard $k$-mers on a one 
dimensional periodic ring and  investigate  the possibility  of phase separation transition in such systems. 
The dynamics consists of a size dependent  directional drive   and  reconstitution  of $k$-mers. 
The reconstitution  dynamics constrained to occur among  consecutive  immobile $k$-mers  allows  them 
to change their size  while keeping  the   total  number of $k$-mers  and the volume  occupied by them conserved.   
We show by mapping   the   model   to a two species   misanthrope process
that its  steady state   has a factorized form.  Along with a fluid phase, the  interplay of drift and reconstitution   
can generate   a macroscopic  $k$-mer,  or a  slow moving   $k$-mer    with a macroscopic  void in front of it, or both. 
We demonstrate   this phenomenon for some specific  choice of drift and reconstitution rates   and  provide  exact   
phase boundaries   which separate the four phases.  
\end{abstract}
\maketitle
\section{ Introduction}
Phase separation in non-equilibrium driven diffusive systems  has been a topic of considerable interest over a long time \cite{zia_book,mukamel_book}. 
Unlike their equilibrium counterparts, such systems have been found to exhibit phase separation, even in one dimension. 
Some well known  examples include 
ABC model \cite{ABC_model}, the LR model \cite{LR_model}, and the EKLS model \cite{EKLS}. A salient feature of these systems has been the presence of  an effective  long-range interaction  or a  short-range dynamics  with   more than one conserved quantities 
being driven through the systems.

A general criterion for phase separation in such systems has been 
conjectured by mapping the driven diffusive system to a zero-range process (ZRP) \cite{psep_criterion}. It was argued that a 
condensation transition in the mapped ZRP would correspond to a phase separation in the lattice. The conjecture was then successfully used to investigate 
the existence of phase separation in AHR model \cite{AHR_model} and later in  the EKLS model.

In all these  driven  diffusive  systems,  hardcore point particles on  a lattice  evolve following a  
dynamical rule  which   might be local, but  effectively generates  long range interaction 
between    particles    due to  the  choice of    hop rates   or   the  presence of   a nonzero
current. Presence  of  extended     objects  could change the  scenario,  as  they bring in   their own  
natural   length scales.  In particular,  reconstitution, if present,  can  generate    extended objects  of arbitrary   lengths, 
 facilitating  the  possibility  of   phase separation.

 The statistical mechanics of diffusion of  extended  objects  has been studied long ago by Tonks in \cite{tonks} 
 to find out the equation of state of the system  in one- and higher dimensions. In one dimension, the  extended  objects   
 with  integer lengths are modeled by   $k$-mers, which 
is an object occupying ${\it k}$  lattice sites and obeys hard-core exclusion. A driven system involving single species ${\it k}$-mer was 
first studied in context of protein synthesis in 
prokaryotic cells \cite{gibbs_protein} to understand the underlying physical mechanism. In a relatively newer study \cite{sasamoto}, the time evolution of the 
conditional probabilities of the site occupation, starting from a known initial configuration has been calculated for such a system. 
Phase diagram and currents in different phases are also  worked out  using an extremal principle 
based on a domain wall theory \cite{lee_protein}. 
Local density evolution governed by the hydrodynamic equations have been framed by mapping the model  of 
diffusing ${\it k}$-mers to  the ZRP \cite{schutz1}.
Other works include studying the effect of local inhomogeneities and defects in  these models \cite{kolomeisky_inhomogeneity, shaw_defect, dong_defect}.
The   spatial correlation functions in  models  with  diffusing  $k$-mers   show 
oscillatory   behaviour which decays  for  large   distance \cite{gupta_kmer};  in the
continuum limit, the  exact  scaling  functions  in such a model are the same as obtained  for a 
 driven Tonks gas \cite{2pt_tonks}

Reconstitution dynamics has been studied   for  one dimensional systems consisting of     only monomers  and dimers 
\cite{barma_DRD,dhar_DRD};  these  systems generally show   strong   ergodicity
breaking.  Later  studies \cite{barma_grynberg,grynberg}  generalized these models  to  include 
asymmetric diffusion  of $k$-mers    up to a maximum length $k_{max}$. This brings  into picture 
kinematic waves  and KPZ non-linearity in general. These models have large number of 
conserved  quantities, {\it i.e.,}  conservation of  number of $k$-mers of all lengths; therefore 
inhibiting  the  possibility of having large  objects. {\clr  In another work \cite{Rakos},  an   asymmetric fragmentation 
process  was studied  where  $k$-mers  could  break    or  combine   when separated by one vacancy; 
for  specific   rates    the model   could    mapped to  the totally  asymmetric 
exclusion process \cite{TASEP}, resulting in  an exponential distribution  of  $k$-mers    with a   
finite cut-off.} 
 
In this  article  we consider a system of hard  poly-dispersed  $k$-mers   undergoing  
asymmetric diffusion and  a  reconstitution dynamics in one dimensional periodic lattice.  The latter allows 
formation of  polymers of arbitrary lengths,  keeping 
 only two conserved quantities,  namely, the total number of  $k$-mers and the  site occupation 
 density. The 
 systems we study  are poly-dispersed, {\it i.e.,} in general the  lattice under consideration  has a mixture of  particles of various sizes. 
 In these models, 
 along with  the size  dependent asymmetric diffusion    rates,   
 $k$-mers  (except monomers) also   undergo a  reconstitution  dynamics. Those  
$k$-mers  which  cannot diffuse   due  to hardcore restriction  can  
 transfer a single monomer unit between them  with  a rate   which depends on their lengths. 
 Since monomers do not participate in reconstitution dynamics, the total number 
 of ${\it k}$-mers is conserved. 
 We show that in presence of  such a reconstitution  process    the system 
 undergoes    novel  phase separation  transitions  forming 
 either    a macroscopically large  $k$-mer, or a macroscopic large void or  both.

 The article is  organized as follows. In section II. we define the model and describe how it can be mapped to 
 a two species ZRP. We follow this
  by showing that the steady state has a product measure. In section III. we study the model for a specific case 
  when the drift velocities are distributed like a step function and explore the possibilities of a four phase scenario. 
  In section IV. we show that for a specific
  choice of diffusion and reconstituting rates, the model has similarities with an ordinary ZRP. 
  We further show that the specific choice indeed leads to a novel phase diagram with four distinct phases.
  Finally,  we summarize the results in section V   along with some discussions.

\section{The model}
We consider $M$ number of  $k$-mers, each    having  different   integer length $k_m$   
with $m=1,2,\dots,M,$ distributed on a one dimensional (1D) periodic lattice of
$L$ sites, following hard core  exclusion.  The sites of the lattice are labeled by 
the index $i = 1, 2, . . . ,L.$ A $k$-mer is a hard extended object  which occupies $k$  
consecutive  sites on a lattice, and can be  denoted by a string of  $k$  consecutive $1$s.
(here represented by $1^k$). 
Thus, every configuration of the system can be 
represented by  a binary sequence  with each  site  being identified as a $0$ or $1$  
 denoting  respectively whether the  lattice 
site is occupied  by a $k$-mer or not.   Thus the total number of vacancies ($0$s) in the system is $N$ and the  total 
length of the  $k$-mers  is  $K= \sum_m k_m $  (total number of $1$s).
 We  define free volume (or void density) as $\rho_0=N/L$  and the  $k$-mer  number density 
density as  $\rho=\frac{M}{L}$.  Thus  $L=K+N$ and the packing 
fraction  (fraction of volume  occupied by the  $k$-mers)  is $1-\rho_0=K/L.$

We assume that there is an intrinsic   drive in the system which forces the  particles to move  in one 
direction, say towards right,  with a rate that depends on  size of the $k$-mer 
\begin{equation}
 k_m 0    \xrightarrow{u(k_{m})}  0k_m     ~~\equiv~~   1^{k_m}0   \xrightarrow{u(k_{m})}  0 1^{k_m}.
 \label{eq:dyn1}
\end{equation}

\begin{figure}[h]
 \centering \includegraphics[width=8cm]{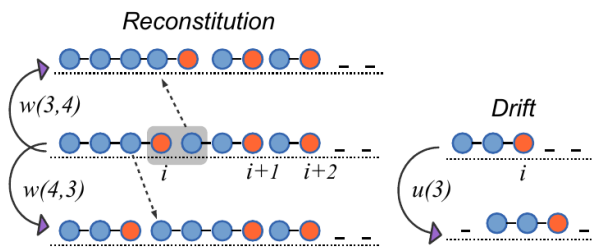}
\caption{Poly-dispersed $k$-mers in  one dimension showing drift and reconstitution. The rate of drift $u(k)$ depends on the length 
of the corresponding $k$-mers.  Reconstitution occurs  only  among  consecutive   immobile $k$-mers
with a  rate $w(k_i,k_{i+1})$  which depends on their lengths. An additional constraint $w(k,k') = 0$ for $k<2$, conserves the 
number of  $k$-mers in the system.
}
\label{fig:dyn}
\end{figure}

Along with  this dynamics  we also  consider a reconstitution  process  among neighboring   $k$-mers 
where  one of the   $k$-mers may release  a single particle (or monomer)  which instantly join the other $k$-mer. 
Reconstitution  occurs  at the interface of  {\it immobile}   $k$-mers which   are expected to  remain 
 in contact for  long time.   
That  is,  as   described  in Fig. \ref{fig:dyn}, reconstitution  process occurs  among two    $k$-mers    $m$ and $m'=m+1$  when they are in contact 
(so that  $m$  is  immobile)  and 
when $m'$ is  blocked  by another $k$-mer  to its right ({\it i.e.} $m'$ is also {\it immobile}):

\begin{equation}
(k_m,~k_{n}) \xrightleftharpoons[w(k_{n}+1, k_m-1)]{\,w(k_m,k_{n})\,} (k_m-1,k_{n}+1).
\label{eq:dyn2}
\end{equation}

The rate   function $w(x,y)$   satisfy $w(1,y) =0,$  {\it i.e.,}  a  $k$-mer     can  release the monomer  only when its size is larger than unity.  
In other words, this restriction  prohibits merging of $k$-mers, thereby   keeping  $M,$ the number of $k$-mers (and thus $\rho$ )conserved.   
It is  evident that  the
dynamics Eqs.   \eqref{eq:dyn1} and  \eqref{eq:dyn2} also  conserves $\rho_0.$

\begin{figure}
\centering \includegraphics[width=8cm]{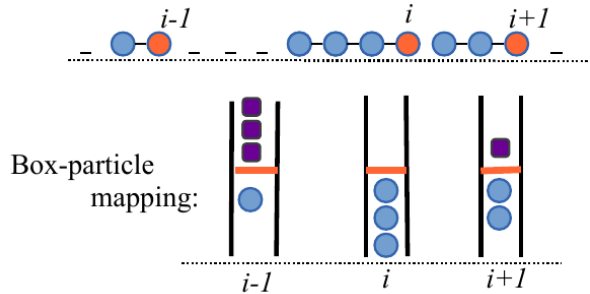}
\caption{ Mapping the drift and reconstitution dynamics of the poly-dispersed $k$-mers to a two species misanthrope 
process (TMAP). The engine
of the $k$-mer is identified as a box and the rest $k-1$ particles as the $k$-particles (circles). 
The number of vacancies following a $k$-mer  are identified as the $0$-particles (squares).}
\label{kmer_map}
 \end{figure}

The dynamics of the model can be mapped  exactly to a  two-species  generalization of  misanthrope process  
 by considering  each $k$-mer  as a box containing  $\m= k-1$ particles  of one kind ($k$-particles)
 and  the number of consecutive vacancies say $n$  in front   of  the $k$-mer   as the  number  of particles of other type ($0$-particles). 
 This is described in Fig. \ref{kmer_map}. Thus  in the box-particle picture,  one of the constituting 
 monomers  of a $k$-mer (say the engine, marked as red in  Fig. \ref{kmer_map})  
 is considered as the  box  containing  two species of particles : additional monomer as 
 one kind of particles  (we refer to them as $k$-particles)  and the   vacancies  in front of  the $k$-mer  as  the  other kind  
 (referred here as  $0$-particles). Thus, we have 
 $M$ boxes containing  $\tilde K =\sum_m^M  \m_m = K-M$  number of  $k$-particles and 
 $N$  number of $0$-particles.  Corresponding particle densities  in  the  two-species misanthrope process (TMAP),   
\begin{eqnarray} \eta= \frac{K-M}{M}=\frac{1-\rho_0}{\rho}-1 ~~~~\text{and}~~~~
 \eta_0= \frac{N}{M}=\frac{\rho_0}{\rho} \cr
\end{eqnarray}
are conserved since   they  can  be expressed in  terms of conserved    densities   $\rho_0$ and $\rho$ defined on 
the lattice. Alternatively the densities   $\eta_0$ and $\eta$  in TMAP uniquely    fixes   the densities on the lattice 
\begin{eqnarray}
\rho_0=   \frac{\eta_0}{ 1+ \eta_0+  \eta } ~~~~~\rho=   \frac{1}{ 1+ \eta_0+  \eta } .
\label{eq:den_trans}
\end{eqnarray}

 To  illustrate the  $k$-mer dynamics  in  the  corresponding  TMAP,  
a  straightforward generalization of misanthrope process to two species, first let us   define   the dynamics  explicitly. 
In TMAP,    $0$- and  $k$-  particles  hop   out  from a box $m,$  
containing  $(n_m, \m_m)$  number of   $0$- and $k$-particles respectively to one of the neighboring boxes $m'= m\pm1$ 
having  $(n_m', \m_m')$ particles  with rates  $u_0( n_m, \m_m;n_{m'}, \m_{m'})$  and   $u_k( n_m, \m_m;n_{m'}, \m_{m'})$.    
Now, the dynamics   of  the   reconstituting  $k$-mers   can  be  written as 
\begin{eqnarray}
 u_0( n_m, \m_m;n_{m'}, \m_{m'})& \equiv& \tilde u(\m_m) =  u(\m_m+1)  \delta_{m', m-1}\cr
 u_k( n_m, \m_m;n_{m'}, \m_{m'}) &\equiv &\tilde w(\m_m, \m_{m'}) =w(\m_m+1, \m_{m'}+1) \delta_{n_m,0}   \delta_{n_{m'},0},
\end{eqnarray}
where the  hop rate of   $0$- particles  $u(\m_m)$  does not depend on    
$n_m,$ the number of   particles of its own kind. The   $\delta$-functions  in the  first equation  
forces  $0$-particles to move towards  left (same as $k$-mers moving to right) and those 
in the  last equation  takes  care of   the  restrictions  that   reconstitution   or monomer exchange occurs  
among neighboring boxes only  when   they  are  devoid of $0$-particles (equivalently, when  $k$-mers are immobile).

The dynamics  of  $0$-particles   is  similar to that of a inhomogeneous ZRP \cite{Inhomo} where  hop rate of particles  are different 
at different sites, but   the  background disorder  here  evolve  continuously  
through reconstitution  process.   In fact a two species  ZRP   with one species following   a dynamics similar to that of $0$-particles  
has been studied previously \cite{2sp_zrp}. 
The  dynamics  of $k$-particles   is   however  different  here,   and it is  similar to the misanthrope process  \cite {misanthrope} where 
hop rate depends on both occupation  number of  departure and target   boxes, but  here  we have additional  interaction   coming from 
the condition that  both boxes must not contain  $0$-particles during reconstitution.   Thus we have   an interacting 
TMAP  where  one species   does misanthrope process in absence of  the  other species. 
The other species, on the other hand,   hops  in one direction with a inhomogeneous rate.
Below we  show that   TMAP  in this case  evolves  to a  factorized 
steady state when  $\tilde w(x,y)$  has a product form.

 A product measure for TMAP  would imply that the steady state probability   of the system 
 in a particular configuration can be written  in the following form:
\small
\begin{equation}
 P\left(\{\m_i,n_i\}\right)=\frac{1}{Q^M_{\tilde K,N}}~~\prod_{i=1}^M f\left(\m_i,n_i\right)\delta \left(\sum_{i=1}^M \m_i-\tilde {K}\right) 
\delta \left(\sum_{i=1}^M n_i- N\right)
\label{eq:prod}
\end{equation}
\normalsize
The $\delta$- functions introduced in the above equation ensures that the particle numbers 
 $\tilde K$  and $N$  are  conserved. Here, 
$f(\kappa,n)$ is  some function yet to  be determined  in terms of the hop rates,  and 
\small
\begin{equation}
 Q_{\tilde K,N}^M=\sum_{\{\m_i\},\{n_i\}}\prod_{i=1}^M f\left(\m_i,n_i\right)\delta \left(\sum_{i=1}^M \m_i-\tilde {K}\right) 
\delta \left(\sum_{i=1}^M n_i- N\right)
\label{eq:Q}
\end{equation}
\normalsize
is the canonical partition function. It  is   clearly  evident that  the
the steady state condition  on the master equation 
can be  satisfied  if  the  $0$-particles satisfy a  {\it pairwise balance condition} \cite{pairwise},
 \begin{eqnarray}
  &&P(\{ \dots ,\m_{i}n_{i},\m_{i+1} n_{i+1}, \dots \})\tilde u(\m_{i+1})\cr
   &&=  P(\{ \dots ,\m_{i}n_{i}+1,\m_{i+1} n_{i+1}-1, \dots \})\tilde u(\m_{i})
   \label{eq:pb}
\end{eqnarray}
and  the $k$-particle  hop rates satisfy   a detailed balance condition 
\begin{eqnarray}
  &&P(\{\m_1n_1, \dots ,\m_i0,\m_{i+1}0, \dots \m_M n_M\})\tilde w(\m_{i},\m_{i+1})\cr
   &&~~=P(\{\m_1n_1, \dots ,(\m_i-1) 0,(\m_{i+1}+1)0, \dots \m_M n_M\}) \cr
   && ~~~~~~~~~~~~~~~~~~~~\times \tilde w(\m_{i+1} + 1,\m_{i}-1) \label{eq:db}
\end{eqnarray}
These two equations along with  the   factorized steady state  ansatz  leads  to
  \begin{eqnarray}
   &&f(\m_i,n_i)f(\m_{i+1},n_{i+1})\tilde u(\m_{i+1})\cr&&~~~~=f\left(\m_i,n_i+1\right)f\left(\m_{i+1},n_{i+1}-1\right)\tilde u(\m_i) 
   \label{eq:pb1}\\
 {\rm and}~~&& f(\m_{i},0)f(\m_{i+1},0)\tilde w(\m_{i},\m_{i+1})\cr
  &&~~~~=f(\m_{i}-1,0)f(k_{i+1}+1,0)\tilde w(\m_{i+1}+1,\m_{i}-1).\label{eq:db1}
 \end{eqnarray}
Equation \eqref{eq:pb1} is quite  similar to the condition  required   to be satisfied in order to obtain factorized 
steady state in 
 ZRP  with site-dependent hop rate. Here, for any specific configuration,  the  profile of $\{\m_i\}$    of $k$-particles  provide a inhomogeneous   background  on which $0$-particles hop. Although the  background  
 disorder evolve with time, we could obtain   simple conditions ( Eqs.  (\ref{eq:pb1}) and (\ref{eq:db1}))  because the dynamics  here restricts simultaneous   hopping  of  $0$- and 
 $k$-particles. We thus  get  the  solution,

\begin{equation}
 f(\m,n)= \left[\prod_{i=1}^n \frac{1}{\tilde u(\m)}\right]f(\m,0) = \left[\frac{1}{\tilde u(\m)}\right]^n f(\m,0),
\label{eq8}
\end{equation}
where  $f(\m,0)$  depends on the  choice of   the reconstitution rate $\tilde w.$
A factorized choice of rates $ \tilde w(\m,\m')=A(\m) B(\m')$   (when  $f(0,0)$  is   set to unity)  gives,
\begin{equation}
f(\m,0)= \frac{B(\m-1)}{A(\m)}f(\m-1,0) = \left[\prod_{\m '=1}^\m \frac{B(\m '-1)}{A(\m ')}\right].
\label{eq:solution}
\end{equation}
Using   Eq. \eqref{eq8}  we  now get, 
\begin{eqnarray}
 f(\m,n) =u(\m)^{-n}\prod_{\m '=1} ^\m \frac{B(\m'-1)}{A(\m')}.
\end{eqnarray}
Note that the functional form of  $f(\m,0)$  in Eq. \ref{eq:solution} is similar to what  has been  obtained for misanthrope process 
when the hop rate has a  product form, $ \tilde w(\m,\m')=A(\m) B(\m')$ \cite{misanthrope}. However, in this  model,  $k$-particles hop 
symmetrically   in {\it  both} directions (though with   different rates) allowing  us to frame a detailed balance condition which does not impose any further  restrictions on the 
functional form of  $A$ and $B.$  For  directed hoping, the simplest possible balance condition 
is the pairwise   balance, which  restricts   $A$ and $B$  to  differ  at best by a  constant : 
$A(x) = B(x) -\alpha.$  We will study the model in presence of these restrictions in  the following 
section and in Sec. IV we explore the possibility of phase separation   transition  when $A$ and $B$ are 
independent functions.

Our  objective  here is  to  investigate  if   reconstituting  $k$-mers  can undergo  a  phase separation transition in   
one dimension.  This is equivalent to the   possibility of having a    condensation transition in  TMAP.    
To this end  we  study the  model  in  the grand canonical ensemble (GCE) and  locate the regions 
in parameter space where   macroscopic densities  has a critical limit beyond which    the densities  cannot be controlled  merely 
by tuning the fugacities. Therefore in  these regions,  a system  having  a conserved  density larger than 
the critical limit   is expected  to produce    a condensate    carrying a finite fraction   of  particles in the system. 
Since there are two  conserved    densities, there  could be four phases  in general.   In   the  TMAP
the  four  distinct phases would  correspond to   condensation of either  of the  two species, or both   or none of them.

The  grand canonical partition function  can be written for  Eq. \eqref{eq:Q} as
\begin{eqnarray}
Z(x,z)&=&\sum_{N=0}^{\infty}\sum_{\tilde{K}=0}^{\infty} x^{\tilde{K}} z^N Q_{\tilde{K},N}^M =F(x,z)^M, 
\label{eq:Z}
\end{eqnarray}
where
 \begin{eqnarray}
  F(x,z)=\sum_{\m=0}^{\infty} \sum_{n=0}^{\infty}x^\m z^n f(\m,n) =   \sum_{\m=0}^{\infty}\frac{f(\m,0) x^\m}{1-z/\tilde u(\m)}
\label{eq:F}
\end{eqnarray}
and $x, z$  are the  fugacities associated with  $k$- and $0$-particles respectively.  
Corresponding densities are then, 
 \begin{eqnarray}
   \eta(x,z)=\frac{x}{F}\frac{\partial F}{\partial x} ~~ {\rm and}~~
   \eta_0(x,z)=\frac{z}{F}\frac{\partial F}{\partial z}
   \label{eq:den}
 \end{eqnarray}
To proceed further  we  need to make a choice of $A(\m), B(\m')$
and the  drift rate $u(\m)$  which   we do in the next section.    

\section{Phase separation transition}

To    illustrate the possible    phase separation transitions  in this model  we    chose a particular 
reconstitution rate

\begin{eqnarray}
w(\m+1, \m'+1)= \tilde  w(\m, \m') =A(\m) B(\m')\cr
~~ {\rm with }~~B(\m)=  \frac{\m+2}{\m+1}  ~~{\rm and}~~A(\m)= \frac{\m+2}{\m+1}-\alpha
\label{eq:misAP}
\end{eqnarray}  
that  produces analytically tractable steady  state  solution.  The parameter   $0\le \alpha \le 1$  here controls the 
reconstitution rate. 
Note  that 
in absence of $0$-particles  ({\it i.e.,} when    the free volume  $\rho_0=0$) the  only dynamics in the system   is  reconstitution of $k$-mers.  The dynamics of 
$k$-particles is similar to  that of  the  misanthrope   process \cite{misanthrope} and   thus   we expect a  condensation 
transition  here.

For $\rho_0 > 0,$  the   $k$-mers also  drift towards right.   Let us  choose   the drift rate  as a step  function
\begin{equation}
 u(k) = \left\{\begin{array}{ll}
          v ~&{\rm for}~ k \le  \Bbbk+1\cr
         ~ 1 &{\rm for}~ k >\Bbbk +1
         \end{array}
\right. \label{eq:uk}
\end{equation} which  has two free parameters $\Bbbk$ and    $v.$  The $k$-mers   having  length  smaller than  a   threshold    $\Bbbk+1$  move    to right with 
 rate  $v$ whereas  others move    with   unit rate.   In    the box-particle   picture,  the  corresponding  hop rate    
 $0$-particles is 
  \begin{equation}
 \tilde u(\m) = \left\{\begin{array}{ll}
          v &{\rm for  }~ \m \le  \Bbbk \cr
          1 &{\rm otherwise}
         \end{array}
\right.
\end{equation}

Thus, the complete dynamics  in the box-particle   picture  can be understood in  the  following way:
$0$-particles  hop to  its  right  on  a   inhomogeneous  background     which  itself evolves  with time 
as $k$-particles  also  hop symmetrically following a misanthrope process  \eqref{eq:misAP}.
In this  case  (up to a constant factor),
 \begin{eqnarray}
 f(\m,0)=\frac{\m!}{(c)_\m}(\m+1)^2, ~{\rm with}~~ c= \frac{3-2 \alpha}{1-\alpha}
\label{eq:fk0}
\end{eqnarray}
where $(c)_\m=c(c+1)\cdots(c+\m-1)$ the Pochhammer symbol.  Thus   the partition function 
in the GCE, following Eqs.  (\ref{eq:Z}) and (\ref{eq:F}),  is   $Z(z,x)=F(x,z)^M,$    with 
 \begin{eqnarray}
 &&F(x,z)=   \frac{z(1-v) G_{ \Bbbk}(x)  +  (v-z) G_\infty(x)}{(1-z) (v-z)}\\
 &&{\rm where }~G_ \Bbbk(x) = \sum_{\m=0}^ \Bbbk   \frac{\m!}{(c)_\m}(\m+1)^2  x^\m.
\end{eqnarray}
It is evident   that   the  maximum value of the fugacity $z$ is  $z_c=  min\{1, v \}$   and that of 
$x$ is  $x_c=1.$ Corresponding densities are then, 
  \begin{eqnarray} 
    \eta_0(x,z)&=& \frac{z}{1-z} +  \frac{ z v(1-v)G_\Bbbk(x)}{(v-z)[z(1-v) G_{ \Bbbk}(x)  +  (v-z) G_\infty(x)] }\cr  
    \eta(x,z) &=& x \frac{z( 1-v) G_{ \Bbbk}'(x)  +  (v-z) G'_\infty(x)}{z ( 1-v) G_{ \Bbbk}(x)  +  (v-z) G_\infty(x)}. \label{eq:rhox}
  \end{eqnarray}
  controlled by     $0<x<x_c$ and   $0<z<z_c.$   Here  $'$  indicates  derivative with respect to    $x.$ 
  Since both the densities are increasing functions of    $x$ and $z$, 
  the maximum  achievable macroscopic  density lies  along the lines  $ x=x_c$ and  $z=z_c.$  Thus  we can have four 
  different phases    (A) the fluid phase ($x<x_c$ and $z<z_c$)     (B) $k$- particle condensate   ($z<z_c$ and $x=x_c$) 
  (C)  $0$-particle condensate  ($z=z_c$ and $x<x_c$)and  (D)  condensation of both  ($x=x_c$ and $z=z_c$).  
  
\begin{figure}[t]
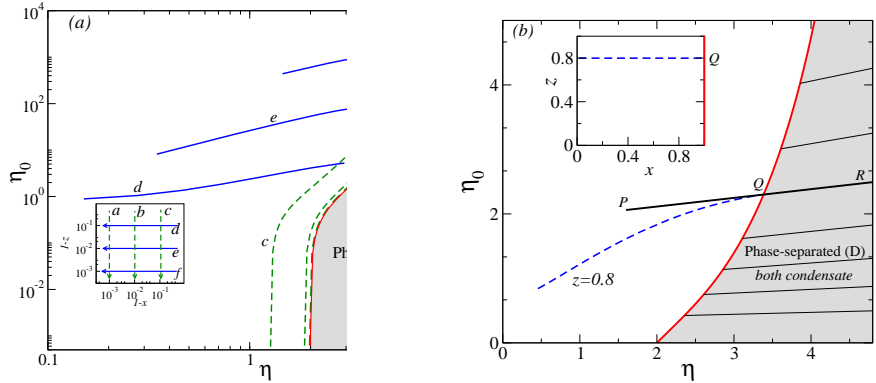

\centering \includegraphics[width=6.cm]{c7v2}\includegraphics[width=5.5cm]{c7v2B}
\caption{  (a)  Log-scale plot of $\eta_0$ vs $\eta$   for some chosen contour  lines  shown in 
$x$- $z$ plane (inset). The  contour lines   $c, b, a$ for   constant $x$  approaches a  limiting curve  (red-line)
when $x\to 1$ whereas  contours  $d, e, f$   for constant $z$  diverges as  $z \to 1.$ 
(b)   The constant $z$  contours, which are normal to  $x=1$ line,    approach  the   critical line 
in $\eta$-$\eta0$ plane   with a   different  slope;  we demonstrate this   for    $z=0.8$ by   drawing 
a  straight line  PQR  at the critical point Q. Systems   with densities   on the line  QR   reach a 
steady state  where background density is   given by  the critical  point Q \cite{Gross08}. 
Since the slope is positive,  in this case  both  $k$- and $0$-particle  will have a condensate.  
For illustration, here we    have taken    $c=7$ and $v=2.$ }
\label{dif_rec_kasep}
 \end{figure}

 One   interesting scenario is  $z=0$  where   $\eta_0 = 0$  and    $\eta(x,0) =  xG'_\infty(x)/G_\infty(x)$  which 
 in the limit $x\to 1$ gives  $\eta^c=4/(c-5).$  This is same as   a   misanthrope process   with   rates given by Eq. \eqref{eq:misAP} $-$  accordingly 
 the partition function is  given  by $G_\infty(x)^M.$ In the lattice model,  however  the free volume   is $\rho_0=0$ and    $k$-mer 
 number density   beyond which      the condensation transition occurs is  $\rho^c =  1/ (1+ \eta^c) = (c-5)/(c-1).$  
 This  transition is  different  from  the usual   phase separation   transitions observed in lattice models.  Here the free volume  being zero, 
 the only dynamics is  the reconstitution process  which   produces {\it one}  macroscopically large  $k$-mer in the pool 
 of  other $k$-mers  of  much smaller sizes.

 Another   interesting limit   is $z\to z_c.$ In this case,  $\eta_0$   in Eq. (\ref{eq:rhox}) diverges
 (as the  first-term  in r.h.s  
 diverges   when $v>1$ and the second term when  $v<1$)    which results  in $\rho_0= \eta_0/(1+  \eta_0 +\eta) \to 1.$  
 Thus,   the critical value density for $\rho_0=1$  is $\rho^c =0.$    In  other words,   when the  $k$-mer    number density is   very low,   they   
 always form a condensate irrespective  of the  value of $v.$

We proceed further  with   $\Bbbk=2$ and   $c=7$  to demonstrate    the  possibility of   phase separation. 
 In this case  monomers ($k=1$ or $\m=0$),  dimers ($\m=1$) and  trimers ($\m=2$)    move   with rate $v$ whereas  
 other $k$-mers     move  with   rate $1.$   Calculations  for  non-integer $c$  or $\Bbbk>2$ are straight forward and  
 generate long   expressions involving  Hypergeometric functions, but it does not provide any additional physical insight.  
 Now, the densities   can be calculated  
from  Eq. \eqref{eq:rhox} by using 
\begin{eqnarray}
  G_\infty(x) &=& \frac{30}{ x^6}   (5-x) (1 - x)^3 Log (1 - x)    + \frac{5}{2 x^5} (2-x)(30 -66 x +37x^2)  ; \cr
   G_2(x) &=&1( 16 x + 9 x^2)/28
\end{eqnarray}
Each point in  $x$-$z$ plane   corresponds to  unique densities  ($\eta_0,\eta$). 
To obtain  the extremal limits of these densities  we plot variation of $\eta_0$ as a function $\eta$  for   
different contour lines in $x$-$z$ plane which  approach   to either $x=x_c=1$  or to $z=z_c=1$.  
Figure  \ref{dif_rec_kasep}(a)  demonstrates, for $v=2$, how 
$\eta_0$  varies   with  $\eta$   when  the fugacity $x$ approaches its critical limit $x_c$   following  the 
the contours $x_c-x=0.1, 0.01$ and $0.001$  (denoted as   solid  lines  $c, b,a$). 
Similarly the contours  of  $z_c-z=0.1,.01, 0.001$  and corresponding  density variations in $\eta$-$\eta_0$ plane 
are shown  as  dashed lines  $d,e,f$.  
It is evident that when   $z$ approaches its critical value $z_c=1$ and $0 \leq x \leq x_c,$  the lines   in  $\eta$-$\eta_0$ plane  
diverges.  However, for  the  other limit   $x\to 1$  (with  $0 \leq z \leq 1$)  the   contour lines   in  $\eta$-$\eta_0$ plane reaches a  
limiting  curve  and thus  densities   beyond  this line (shown as shaded  region)   cannot  be achieved in the  GCE by tuning the  fugacities 
$x$ and $z.$   The equation  of this  limiting  or critical line,  $x=1$ ,  can be translated to  $\eta$-$\eta_0$ plane 
by expressing $\eta_0^c\equiv \eta_0(1,z)$ as a function of $\eta^c\equiv \eta(1,z)$  (red lines in   Fig.  \ref{dif_rec_kasep}(a)  and (b)).

Thus the  density conserving  dynamics, given by   Eqs. (\ref{eq:misAP}) and (\ref{eq:uk}), cannot  distribute   particles   
macroscopically if densities  $(\eta, \eta_0)$   lie in this region.   But what would be the   background density ?   

In a  recent work  \cite{Gross08},  Gro\ss kinsky     proposed a  procedure  to obtain the  background density  from  the grand 
canonical distributions.  It was argued that    the  normal  directions   of  the critical line  in   
 $\mu_x$-$\mu_z$  plane (here chemical potentials   are   $\mu_{x} = \ln(x)$ and  $\mu_{z} = \ln(z)$)
 translates to a direction   in    density plane  along which  the  background density  remain invariant. 
  For  TMAP,  the critical line is    $x=1$   ({\it i.e.}   $\mu_x=0$)   and thus the normals are defined by 
 $z=constant.$  For illustration, we  take $z=0.8$  in  Fig.  \ref{dif_rec_kasep}(b)  and  plot the corresponding   contour  
in $\eta$-$\eta0$ plane  (shown as dashed line). It  approaches   to  the  critical point  $Q \equiv (\eta^c ,\eta_0^c)$     
with a   slope  given by   the tangent line  PQR.     The Gro\ss kinsky criteria   indicate that   the  background density 
along the line  QR    in  the condensate phase   is invariant  and  is given by  the critical point Q.  Clearly, for  any arbitrary
density   $(\eta, \eta_0)$  on the   line  QR,   $\eta>\eta^c$  and     $\eta_0>\eta^c_0$;  both  species would have
extra particles   which  would form a condensate.   Such lines,   drawn  in  Fig.  \ref{dif_rec_kasep}(b) for other values of $z$
clearly  indicate   that   both $k$ - and $0$- particles condensate in   phase D.

\begin{figure}[h]
\centering \includegraphics[width=10cm]{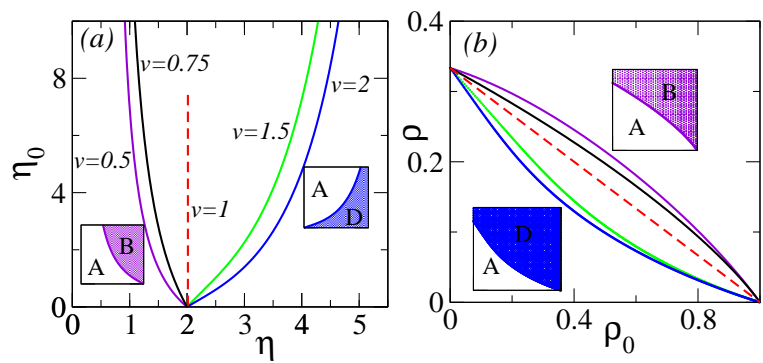} 
\caption{(a) Phase diagram in the $\eta -\eta_0$ plane. The critical lines
corresponds  to $v=0.5, 0.75, 1, 1.5$  and $2$ (from left to right along $\eta$ axis). 
The red-dashed line  $\eta=2$  is  critical   when $v=1$, {\it i.e.,} when    all $k$-mer moves with same  rate. 
Figure (b)  shows  the corresponding transition lines  in $\rho-\rho_0$ plane, translated from $\eta$-$\eta_0$ using Eq. (\ref{eq:den_trans}). 
The shaded regions in the insets represents  the  phase separated state B (only $k$-condensate)   when $v<1$  and 
D (both  particles condensate)   when $v>1.$}
\label{fig:c7v}
 \end{figure}

 For    generic values of $v>0$   the  densities $(\eta, \eta_0)$   are finite along the line $x=1$ and 
the  critical line  $x=1$    translates to,  

\begin{eqnarray}
\eta^c&=& 2\frac{(53  +17 v) z -70 v}{ (53v +17) z -70v} ~~~~~ or~~~ ~ z= \frac{70v(\eta^c-2)}{ 17 (\eta_c - 2v)  + 53(\eta^c v -2)}\cr
\eta_0^c&=& \frac{z}{1-z}  +  \frac{v}{v-z} +\frac{70 v}{(17+53 v) z - 70v } \cr
  &=& \frac{(2-\eta^c)\{ 17 v (53 \eta^c-34)^2 +53 (17 \eta^c-106)^2 \}}{72 (v-1) (53 \eta^c-34)(17 \eta^c-106)}
 \label{eq:critLine}
\end{eqnarray}
\vspace*{.2 cm}
The critical lines   for the condensation transition   in TMAP   are   shown in Fig. \ref{fig:c7v}(a) 
for different values of $v.$   In the inset  we   show the  $k$-particle condensate phase (shaded  region)  for  for $v=1/2$ and  $v=2.$

Note  that   for $v<1$ the  condensate phase    is of type B where only $k$-particle  condensate.  
Unlike $v>1$ case,  here $0$- particles  hop  out  faster  from    the  sites with large  $k$ values   (particularly the condensate site)
inhibiting     formation of condensation. Again, for $v=1,$
  all $k$-mers   drift with equal rate, and thus  the  reconstitution dynamics  become  identical to 
a misanthrope process  of $k$-particles,  forming  a  condensate   for  $\eta>\eta_c =4/(c-5)=2.$  
Also,  for $\eta_0 =0$ the   condensation always 
occurs  at $\eta =2$ independent   of $v$    as   in this case  there is  no  free volume   available for    drift.  
The critical lines can be explained in the following way: for $0<v \le 1$ the possible values of $\eta^c$  lies  in the range  $0<\eta^c\le 2.$ 
However from Eq. \eqref{eq:critLine} we have a divergence of $\eta_0^c$ at $\eta^c=34/53$. Thus the critical line for $v<1$ is obtained by 
choosing the accessible values for $\eta^c$
in the range $34/53 <\eta^c<2$ and obtaining the corresponding values for $\eta_0^c$ from Eq. \eqref{eq:critLine}. 
Similarly,  to ensure positivity and finiteness in $\eta_0^c,$  the critical line for $v>1$ must  be bounded in the 
region $2 <\eta^c<106/17$.
Note that  when  $\eta_0$ is increased,  the transition occurs  for  lower  $\eta$ values   when   $v<1$, {\it i.e.,}  when 
monomers,  dimers and trimers  move slower than   other $k$-mers. Naturally, this   scenario   reverses   for  $v>1.$
It is  also evident that for   $\eta > 106/17$  condensation occurs  for any $\eta_0>0.$   

To  observe 
the phase separation  transition,  we  finally   translate these transition   lines  in $\eta$ -$\eta_0$  plane to 
 $\rho_0$-$\rho$  plane   in Fig.  \ref{fig:c7v}(b)  using Eq. (\ref{eq:den_trans}). 
$\eta_0^c$  in  Eq. \eqref{eq:critLine} diverges in this limit.  In Fig.  \ref{fig:c7v}(b), the phase separated  
state  is   shown as  shaded region , for  $v=1/2$ and $v=2.$

\section{Four phase  scenario}

 In the previous section we have   observed   only   some of the condensate   phases where  due to reconstitution,   
one of the $k$-mers  may become  macroscopically large and   $0$-particles    are attracted to the condensate site  when 
it moves   with a  slower  rate.    It is however possible to have  an independent  condensation 
of vacancies as  they  experience an  evolving disordered background:  slow moving $k$-mers  may  accumulate   
macroscopic   number of vacant sites  in front of it  forming an additional condensate phase. 
To study all these possible  scenarios  we  choose  a  hop rates  having a  single parameter $b>0$ as
\begin{equation}
A(\m) = \m+b, ~~ B(\m) = \m +1  ~{\rm and }~  \tilde u(\m) = \frac{\m+2}{\m+1}.
\end{equation}
In this case $f(\m,0)=\m!/(1+b)_\m$ and thus 
\begin{equation}
           f(\m,n)=  \frac{\m!}{(1+b)_\m} \left( \frac{\m+1}{\m+2}\right)^n \label {eq:fmn}
    \end{equation}
Note that  the weight factor  $f(\m,0)$  is  same as what one  gets for an ordinary ZRP \cite{ZRPrev} with particle hop rate $u_{_{ZRP}}(k) = 1+ b/k$ 
(even though  $k$-particles  here  follow a misanthrope dynamics).  Thus,  in absence of $0$-particles  it is  assured  that  a condensation transition  occurs for 
large densities when $b\ge2.$    In absence  of $k$-particles , however, $0$-particles follow a simple ZRP dynamics  with a constant rate 
$\tilde u(0)=2$ and they cannot   form a  condensate.  We  proceed  to see   that the interaction between  $0$- and $k$-particles   
can   provide an  additional phase   where    {\it only}  $0$-particles   condensate.

\begin{figure}[h]
 \centering \includegraphics[width=12cm]{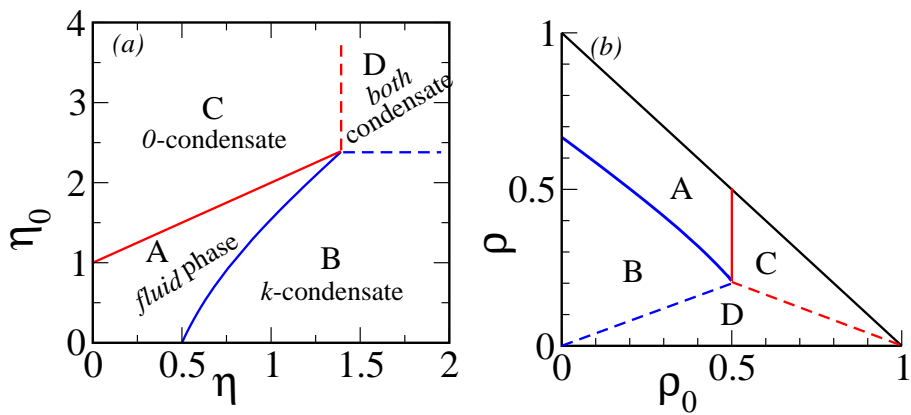}  
\caption{In (a), the four phases in the $\eta_0-\eta$ plane is shown for $b=3$. In (b), the corresponding phase diagram 
 in the original lattice variables is shown.}
\label{fig:4pb3}
 \end{figure}

 A  model with   steady state  weight   given by  Eq.  (\ref{eq:fmn})   has been  studied
 in context of two-species   ZRP \cite{2sp_zrp}    which   exhibit all  possible  condensate  phases   depending on    the value of 
 $b$  \cite {2sp_zrp, Gross08}.   For completeness we revisit     the  model     for $b=4$     where   all  different condensate 
 phases are present; we then  translate  the  transition lines  to obtain the    phase separation    scenario 
 in    driven $k$-mer   models    in one dimension.

In GCE the partition function is $Z(x,z) = F(x,z)^M$.  Following Eq. \eqref{eq:F}  we  get
 \begin{eqnarray} 
 F(x,z)&=& \left[36 x^4(2z-1)(3z-2)\right]^{-1}~ [ ~x^2 (19 x z-14 x +42 z-30) \cr
  &-& 6\ln(1-x) \left\{  x (3z-2)(2z x^2-x^2-3) +2(2z-1) \right\} \cr
  &-&12 (2 z-1)  - 36 x^4(1-z)^2  \Phi \left(x,1,(2-z)/(1-z)\right) ~ ]
 \end{eqnarray}
 
 where   $\Phi(.,.,.)$ is  the Lerch Phi  function, and   factors  which are independent of $x$ and $z$ are ignored.   
The corresponding densities can  now be  obtained using 
Eq. (\ref{eq:den}). We avoid writing down those long expressions explicitly. Instead, in the following, we focus on the 
results   only at the critical limits. 
It is evident that  $F(x,z)$ diverges as $x \to x_c=1$  (logarithmically) and  for $z \to z_c=1$.

As long as  the        equivalence of grand canonical  and   canonical  ensemble  holds
{\it i.e.,} in the density regions bounded by  the parametric  curves  $\{ \eta(x,z_c), \eta_0(x,z_c)\}$ and  
$\{ \eta(x_c,z), \eta_0(x_c,z)\}$  in the    $\eta-\eta_0$ plane, the system remains in the  fluid-phase. 
Condensation occurs when these parametric    functions   become finite valued  $-$ then  in the canonical ensemble, 
densities  larger than these  values  does  not have any representative fugacities, thus resulting in condensation. 
{\clr 
For $b=4$  we have  the critical lines, 
\begin{equation}
 \eta(x,1)=\frac{x(11 x^2+30x-48)-6 \ln(1-x)(8-9x+x^3)}{x(12-12x-5x^2)+6 \ln(1-x)(2-3x+x^3)}
\end{equation}
\begin{equation}
\eta_0(x,1)=1+\eta(x,1)
 \end{equation}
\begin{equation}
 \eta(1,z)=h(z)   \left[ 74 - 67z-36(2-z)(1-z) H_{\frac{1}{1-z}}\right]
\end{equation}
\begin{eqnarray}
 \eta_0(1,z)&=& 6 z h(z) \left[ 6 \psi_1\left(\frac{2-z}{1-z}\right) \right.\cr 
 ~~~~&&+ \left. \frac{  64z -37z^2-25+6(1-z)(3-5z)\{\gamma+\psi_0\left(\frac{2-z}{1-z}\right)\}}
 {(3z-2)(2z-1) }\right]
\end{eqnarray}
 where  $h(z)= [36(1-z)^2 H_{\frac{1}{1-z}}+37 z-32]^{-1}$   with  $H_m$  being  the Harmonic    number,   
 $\psi_m(.)$ is    the poly-gamma function of    order $m$ 
and $\gamma$ is the   Euler constant.
}

\begin{figure}[h]\centering \includegraphics[width=10cm,height=9cm]{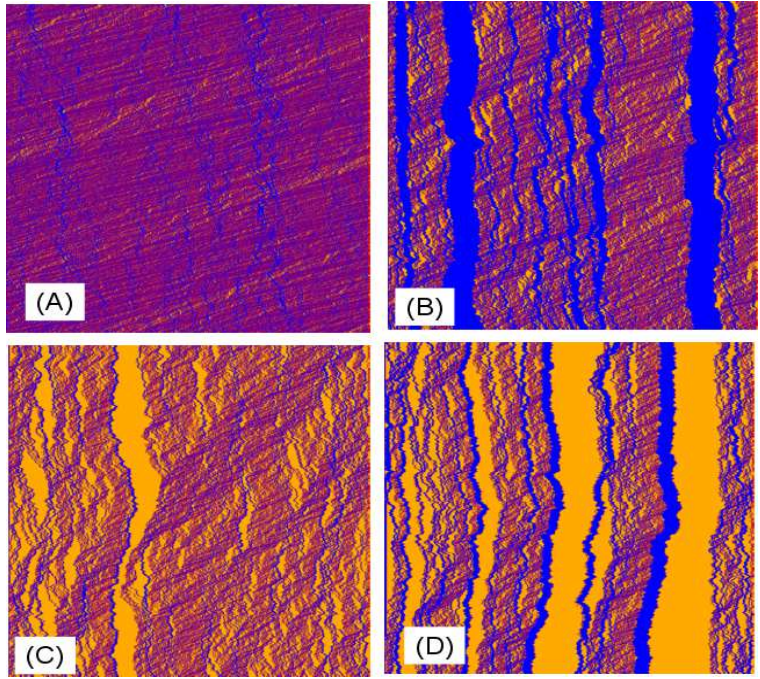}
\caption{Space time plot  of $k$-mers after   $10^5$  MCS. Four different phases A (fluid), B ($k$-condensate), 
C ($0$-condensate, and D (condensation of  both $k$- and $0$-particles
are   shown here for  densities $(\rho_0, \rho)= (1/6, 2/3),(1/4, 1/4), (5/8,1/4)$ and $(5/8,1/8)$ 
respectively. $k$-mers are    represented by $k-1$ blue pixels ending with a  red pixel corresponding 
to the engine,  vacancies are colored as orange. Here, $L=900.$}
\label{fig:t_evo}
 \end{figure}
In Fig. \ref{fig:4pb3} (a)  we have   plotted    these  parametric functions  $\eta_0 $ versus $\eta,$  
for $x=x_c$  varying  $0<z<1$       (blue curve) and then  for  $z=z_c$     varying $0<x<1$ (red curve). 
These two critical lines   encapsulate a region    where   macroscopic density is  well described by   
the fugacities $x$ and $z$
(denoted by  A, the  fluid phase). {\clr To   investigate     the  possibility  of  condensation   in the
other  density regions, we    follow the     Gro\ss kinsky  criteria  \cite{Gross08}  described    in the previous section. 
In fact, the  phase diagram  for   a different model  having the same steady state  structure  has been  explored   
in details \cite{Gross08}. }
The $k$-condensate occurs  in the region beyond  $x=x_c$ curve (denoted by B), 
whereas $0$-particles  condensate    in the density  region  above  $z=z_c$ curve in  $\eta $-$\eta_0$ 
plane (denoted as C). Finally the density region    $D$ corresponds to  condensation of both  kind of particles. 
The transition lines  in  box-particle picture  are then translated to the  $k$-mer densities   on the lattice   using 
Eq. \eqref{eq:den_trans}.  Correspondingly,    the  $k$-mers       show   three different kind   of phase separated states  along 
with the fluid phase A, as shown in  Fig. \ref{fig:4pb3}(b).  Note that, the fluid phase  here  corresponds to a phase  
where $k$-mers  are well mixed with the     vacancies. This  requires small
free volume  $\rho_0 \approx 1$  and large   $k$-mer  number density $\rho$ so that size of each $k$-mer is  small. 

To demonstrate the four-phase scenario, we simulate the drifting reconstituting $k$-mer  dynamics on a system of  size $L=900$. 
Starting from a random  initial  distribution of $k$-mers   the system is 
allowed to evolve  for $10^5$ MCS; for the next $10^3$ MCS, we store the   location and size of $k$-mers  and 
color them as follows,  $k$-mer: blue, engine: red and  vacancy: orange. 
Figure  \ref{fig:t_evo} (A) (B) (C) and (D)  shows  the space-time (downwards) plot of these configurations  for four different  
 points $(\eta,\eta_0) = (1/4,1/4), (2,1), (1/2,5/2)$ and $(2, 5)$,
representing  four different  phases respectively.     All three  different condensate  phases   
show striking co-existence features whereas the fluid  phase is well mixed. Note that $k$-particles show a localized condensate as  
reconstitution (or $k$-particle  hop in TMAP) occurs symmetrically, whereas the $0$-condensate  moves  towards left. 

 \section{Conclusion}
 
 In  this article we  study the possibility of phase separation transition in a one dimensional 
 system of  
 poly-dispersed  hard  extended objects  called  $k$-mers   which  occupy 
 $k$  consecutive lattice   sites.   A  $k$-mer   can move  
 by one lattice site  with a  rate  $u(k)$   which depends on  its size,  if the rightward 
 neighbor is vacant.   Along with the drift,   $k$-mers  also undergo reconstitution,  where 
two consecutive $k$-mers  of size $k$ and $k'$, if immobile  due  to hardcore constraint ({\it i.e.}, 
their immediate right neighbor are not vacant) can exchange  one particle   between them 
 with rate  $w(k,k')$; thus their  sizes  become either $(k-1, k'+1)$   or $(k+1, k'-1).$
 This  dynamics conserves the total  volume  occupied  by  the $k$-mers.
 We also  impose an additional conservation  of  the number  of $k$-mers by   restricting 
 $w(k,k')=0$  for $k<2$.  We  find that  this additional conservation   can  lead 
to  four    distinct phases :  (A) a fluid phase where   similar  size $k$-mers 
are well mixed  with the vacancies, (B) one  macroscopic $k$-mer  and   rest of the  system 
remaining homogeneous, (C)  a  very slow moving $k$-mer  leaving a  trail of vacancies (macroscopic void) 
 in front of it, and (D) a macroscopic void  along with a macroscopic  $k$-mer. 

We could obtain the exact boundaries    which separate   four  different phases  by  mapping the  model  
to a  two-species misanthrope process (TMAP)  by  considering   $k$-mers  as boxes 
 containing $k-1$  particles   of one kind (called $k$-particles)  and the  number of   vacancies  
 in front of it  as  particles  of other type (called $0$-particles).  The  effective   dynamics  in TMAP  
 becomes  the  following:   $k$-particles  in boxes which are devoid of $0$- particles  evolve following 
the dynamics of a misanthrope process with rate $w(k-1, k'-1)$  and the   $0$-particles   hop  with a 
 inhomogeneous   rate  that  depend  on   the   number of  $k$-particles  present in the box. 
The interplay   of interaction  between   two different species is now evident.   $k$- particles hop  only when  
both departure and arrival  boxes are  devoid of $0$-particles,  and $0$-particle hop rate depends on the number of  
$k$-particles.   

{\clr We must mention that in  
ZRP and related models   with one  or more species,  
if   a dynamics  involving  asymmetric hopping  of  particles (say, only to rightward  neighbour)  
evolve  to  a  factorized steady state,  then  the 
model  also    shares  the same steady state if   particle  hopping were symmetric.   In    the model that  we have studied, 
in the presence of  directional drive,  reconstitution 
occurs  when two   $k$-mers  are (a)  in contact,  and  (b) the right  $k$-mer  has no empty site to its right. 
In fact,  condition (b)  naturally decouples  the dynamics of $0$ and $k$-particles and    help in  obtaining   
an  explicit  factorized steady state. {\clr  When $k$-mers  drift   only towards right, the  condition (b) can be interpreted as 
follows: only  immobile  $k$-mers  (which  are  more likely to remain in contact)   can exchange particle 
among  them.  In case of  symmetric   diffusion  of $k$-particle, such an interpretation  would  be lost
in $k$-mer picture, but  the steady state  however is  {\it not}  altered. }
To  model  a   more   realistic situation  like  dynamics of polymerization \cite{polymer}  or   
filaments in micro channels  \cite{microchannel},   we need  to   extend this   dynamics  to  higher dimensions and  to  
include   a  fusion dynamics  which breaks   the  conservation   of number of  $k$-mers and a  fragmentation  dynamics 
which   allow    breaking of  a  $k$-mer  into two random segments. }

 {\it Acknowledgments:} {\clr  We thank the anonymous referee  for  his valuable comments  and constructive suggestions. }  
 BD gratefully acknowledges the financial support of Council of Scientific Research, India (Ref. 09/489(0089)/2011-EMR-I). 
 PKM gratefully acknowledges the support of CEFIPRA under Project 4604-3.

\end{document}